\documentclass[a4paper,fleqn]{cas-dc}

\usepackage[numbers]{natbib}

\def\tsc#1{\csdef{#1}{\textsc{\lowercase{#1}}\xspace}}
\tsc{WGM}
\tsc{QE}
\tsc{EP}
\tsc{PMS}
\tsc{BEC}
\tsc{DE}
\usepackage{algorithm}
\usepackage{algpseudocode}
\usepackage{amsmath}
\usepackage{amssymb}
\usepackage{booktabs}

\begin{document}
\let\WriteBookmarks\relax
\def\floatpagepagefraction{1}
\def\textpagefraction{.001}
\shorttitle{UVTran}
\author[1]{Junfeng Zhang}
\ead{12321235@zju.edu.cn}

\title [mode = title]{ UVTran: Accurate Hole-Filling Parameterization  with Transformers}

\begin{abstract}
In industrial design, N-sided hole filling is typically formulated as the construction of a single trimmed B-spline surface by minimizing a fairness energy subject to geometric boundary constraints. This formulation requires an accurate parameter-space representation of the trimming curve on the filling surface. Most existing methods project the hole boundary onto a nearby plane or polygon to establish correspondence; however, they often neglect boundary heterogeneity, which can yield biased mappings, degrade fairness, and even cause filling failures. We propose UVTran, a transformer-based framework that predicts an auxiliary projection surface better to capture the geometric characteristics of the hole boundary. Exploiting B-spline locality, we design a cross-attention mechanism that biases each surface control point toward the nearby hole boundary, preserving local geometric detail. We voxelize control-point coordinates and formulate the fitting problem as a classification task, which reduces the model's sensitivity to small numerical perturbations and noise. We adopt a progressive-resolution training strategy that injects controlled discretization errors at coarse resolutions to mimic distribution shifts, thereby mitigating overfitting and improving generalization at high resolution. On our benchmark, UVTran outperforms both industrial and academic baselines: the tolerance-satisfaction rate improves by $12\%$, and it consistently produces fair filled surfaces even under complex hole boundary conditions. These results suggest that UVTran yields more faithful correspondences and fairer trimmed surfaces across a wide range of N-sided holes. Code is available at  \url{https://github.com/UVTran/UVTran}.
\end{abstract}

\begin{keywords}
Hole filling \sep Parametric curve \sep Deep learning \sep Geometric modeling  \sep 
\end{keywords}

\maketitle

\section{Introduction}
\label{sec:intro}

3D models serve as the primary digital representation of products throughout their development lifecycle. During model construction, routine operations such as fillet transitions, structural repair, and feature deletion often produce 3D N-sided holes with irregular boundaries. Another critical application of N-sided hole filling is to repair N-sided regions induced by extraordinary vertices in Catmull–Clark (CC) subdivision~\cite{Catmull98}. CC subdivision refines an arbitrary mesh into a smooth surface; however, in the presence of extraordinary vertices (i.e., vertices with valence $\neq 4$ ), the resulting surface cannot be represented as a single tensor-product B-spline surface. In such cases, N-sided holes arise around these regions and must be filled to obtain a complete, CAD-compatible surface representation.

\begin{figure}
  \centering
\includegraphics[width=0.5\textwidth]{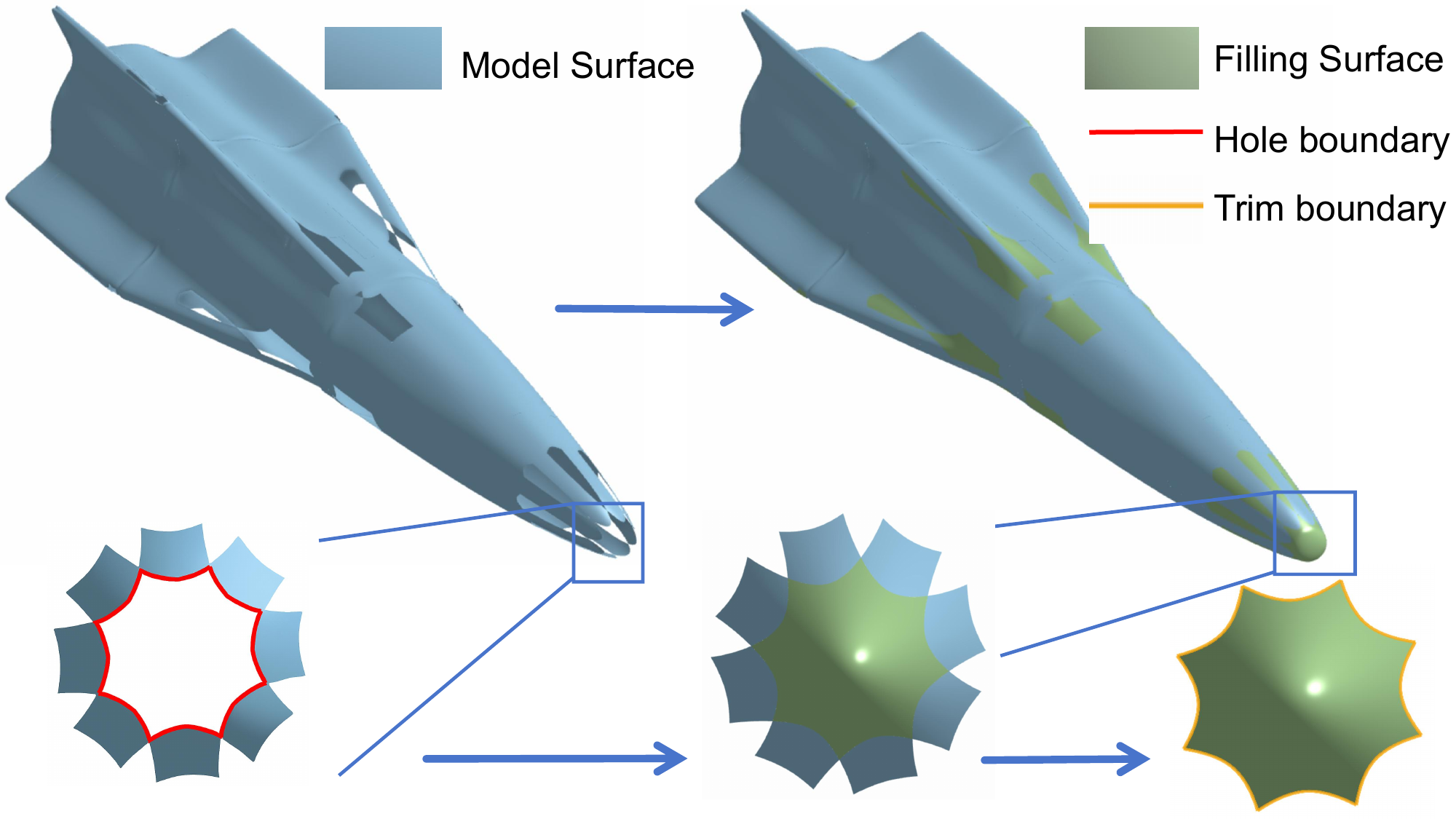}
     \caption{\textbf{Hole-filling results.} Left: original rocket model with multiple holes; right: the hole-filling result produced by UVTran.}
\label{fig:teaser}
\end{figure}

Hole filling aims to generate a surface patch that satisfies prescribed conditions over the hole region. The filled surface must satisfy geometric continuity at the hole boundary—i.e., the errors in position ($G^0$), normal ($G^1$), and curvature ($G^2$) remain within a prescribed tolerance~\cite{farin2014curves}—and must also exhibit surface fairness, as illustrated in Fig.~\ref{fig:teaser}.
Geometric continuity helps prevent stress concentration arising from surface discontinuities. Surface fairness requires that the filled surface be free of local wrinkles and bulges, thereby preserving aesthetic consistency and, when relevant, aerodynamic performance.
Existing hole-filling approaches include trimmed-surface methods~\cite{Liu12,parasolid}, multi-patch constructions~\cite{coons67,Piegl99,Gregory74}, subdivision surfaces~\cite{Catmull98,Levin99}, and generalized surfaces~\cite{Varady16,Vaitkus21}. The use of multiple B-spline surfaces will introduce new internal errors and lead to conflicts at junctions. Subdivision surfaces achieve the required continuity through iterative subdivision, which inevitably increases computational cost and complexity. Although generalized tensor product surfaces can effectively satisfy boundary constraints, operations such as trimming and intersection are extremely challenging, which restricts their direct application in CAD systems. Therefore, trimmed B-spline surfaces are widely used in industry (e.g., Parasolid~\cite{parasolid}) due to their strong local controllability and high compatibility with existing CAD kernels. Using a single trimmed surface for N-sided hole filling requires the trimming boundary of the filling surface to match the input hole boundary up to a prescribed level of geometric continuity. In practice, this requirement can be formulated as a tolerance-bounded discrepancy between the geometric quantities evaluated along the two boundaries, as follows:
\begin{equation}
\begin{split}
\label{geo}
\!\left(\left\|B^{\mathrm{geo}}_{\mathrm{trim}}(t)-B^{\mathrm{geo}}_{\mathrm{hole}}(t)\right\|^2\right)<T^{\mathrm{geo}},\\
\mathrm{geo}\in\{\mathrm{position},\mathrm{normal},\mathrm{curvature}\}, \\
B^{\mathrm{geo}}_{\mathrm{trim}}(t)=S^{\mathrm{geo}}(\mathrm{pcurve}(t)),\\
\mathrm{pcurve}(t)=(u(t),v(t)).
\end{split}
\end{equation}
Here, $B^{\mathrm{geo}}_{\mathrm{trim}}(t)$ denotes the geometric quantities (position, unit normal, and curvature) evaluated along the trimming boundary of the filled surface (Fig.~\ref{fig:teaser}). The $\mathrm{pcurve}(t)$ denotes the corresponding parametric trimming curve(pcurve) in the $(u,v)$ domain of the surface. $B^{\mathrm{geo}}_{\mathrm{hole}}(t)$ denotes the geometric quantities evaluated along the input hole boundary, and $T^{\mathrm{geo}}$ is a user-defined tolerance. 

The key challenge here is that ensuring the geometric continuity between the trimming boundary and hole boundary fundamentally relies on accurately determining the parametric curve $\mathrm{pcurve}(t)$. Because $B^{\mathrm{geo}}_{\mathrm{trim}}(t)$ is obtained via the surface mapping $S(\cdot)$ at parameter locations $(u(t),v(t))$, satisfying Eq.~\eqref{geo} fundamentally depends on determining an appropriate $\mathrm{pcurve}(t)$. In other words, even with a high-quality surface representation, an inaccurate boundary parameterization can prevent the trimming boundary from matching the hole boundary in terms of both geometric and differential properties.

\begin{figure}
  \centering
\includegraphics[width=0.5\textwidth]{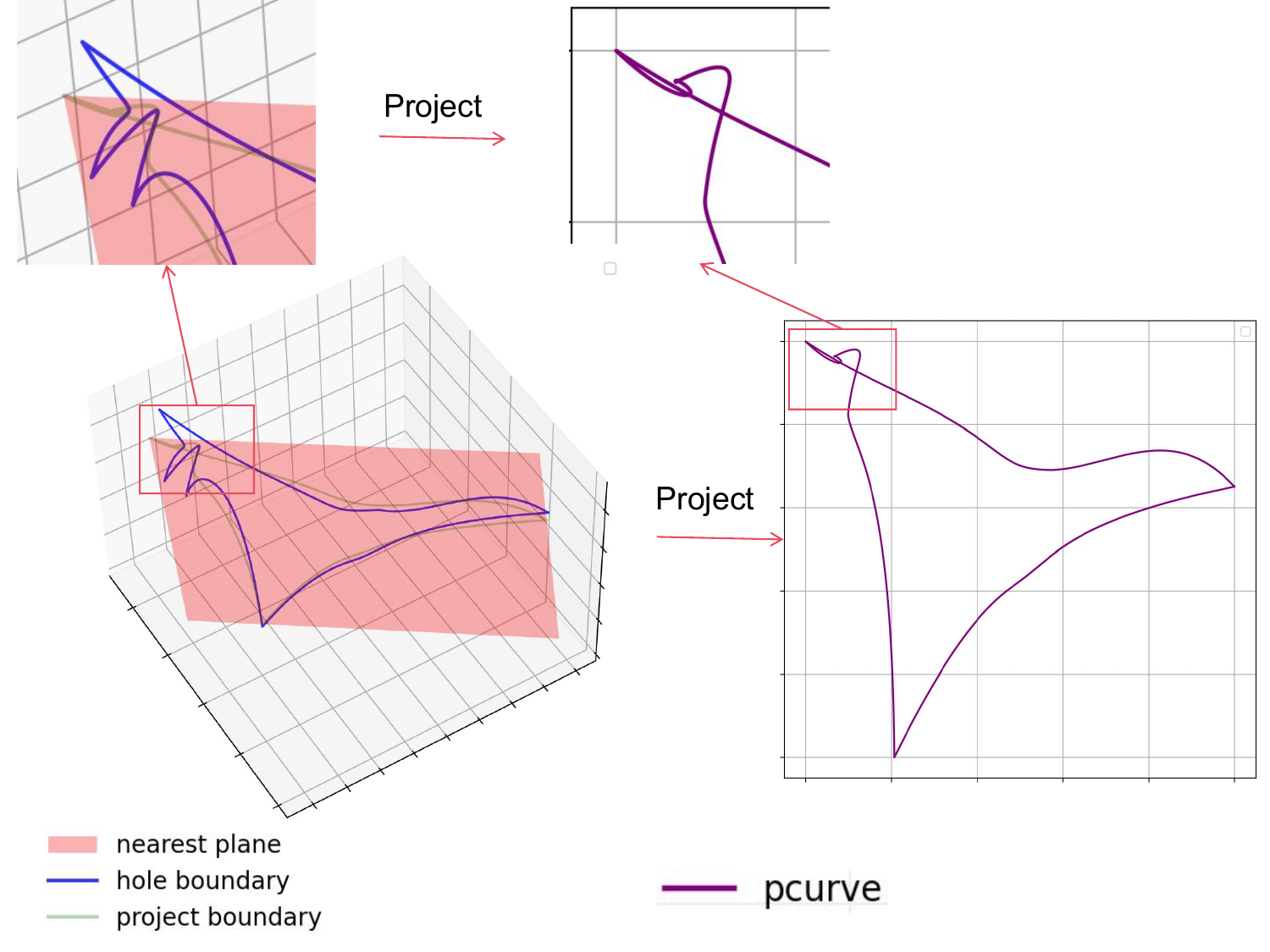}
\caption{\textbf{Self-intersecting pcurve.} NP parameterization can generate self-intersecting pcurve in high-curvature regions.}
\label{fig:baduv}
\end{figure}

Traditional methods for computing 
pcurve include nearest-plane(NP) projections~\cite{Liu12} and mean value coordinates (MVC)~\cite{floater2003}. These methods parameterize the hole boundary by projecting it onto a planar or quasi-planar domain. Although effective for relatively regular boundaries, these methods often treat the boundary as globally homogeneous and therefore adapt poorly to local geometric heterogeneity, such as pronounced concavities/convexities, non-uniform sampling density, or localized high curvature. Consequently, the parameter distribution along $\mathrm{pcurve}$ becomes imbalanced: small geometric features may be over-compressed in parameter space, whereas flatter segments may be over-stretched.  In some cases, the pcurve may even self-intersect. as illustrated in Fig~\ref{fig:baduv}.  This distortion propagates to the trimming boundary $B_{\mathrm{trim}}(t)$, yielding a mismatch with $B_{\mathrm{hole}}(t)$ that cannot be eliminated without significant deformation of the surface interior. In practice, enforcing Eq.~\eqref{geo} under a distorted $\mathrm{pcurve}$ may require sacrificing surface fairness and can still lead to undesirable outcomes, including unfair filled surfaces, self-intersections, or even failure to close the hole robustly.

To address these limitations, we propose UVTran, a supervised Transformer-based framework~\cite{vaswani2017} that learns a boundary-adaptive parameterization via an intermediate projection surface. The central idea is to avoid prescribing a fixed planar domain a priori. Instead, UVTran predicts a projection surface that is strongly correlated with the local geometry of the hole boundary. To obtain the parametric curve, we project the hole boundary curves onto the model-predicted initial surface to obtain projected spatial curves. The corresponding parametric representation of these curves on the initial surface is used as the target parametric curves $\mathrm{pcurve}$. So that the resulting $(u(t),v(t))$ coordinates inherit the boundary’s geometric characteristics more faithfully and in a spatially adaptive manner. UVTran uses cross-attention to explicitly couple boundary features with the predicted surface representation, enabling the projection surface to reflect local shape variations rather than only global trends.
To improve robustness and reduce overfitting, we further introduce (i) a voxelization-based representation and (ii) a progressive coarse-to-fine generation strategy. First, the control points of the target projection surface are discretized into voxel indices, reformulating surface fitting as a classification task. This discretization reduces the model's sensitivity to small perturbations in the training data and alleviates regression instability caused by noise or outliers. Second, the model predicts voxelized control points progressively from coarse to fine resolutions. The coarse stage captures global structure and provides a regularizing prior; the subsequent refinement stage incorporates boundary information to recover fine-scale details. Importantly, discretization error at the coarse level acts as an implicit regularizer, which empirically reduces overfitting during high-resolution prediction and improves generalization to unseen boundary geometries.

The main contributions are summarized as follows:
\begin{enumerate}
\item \textbf{Transformer-based hole-filling framework (UVTran).} We propose UVTran, which predicts a projection surface that is strongly correlated with the hole boundary geometry via cross-attention. The trimming $\mathrm{pcurve}$ obtained by projecting the boundary onto this surface more accurately reflects local geometric variations, thereby improving the precision and stability of single-trimmed-surface hole filling for complex boundary configurations.

\item \textbf{Per-coordinate voxelization for stable learning.} We voxelize each coordinate component of control points independently, casting the fitting task as a multi-class classification problem without introducing an excessive number of categories. Compared with direct 3D voxelization at the exact resolution, this design substantially reduces the label space, lowers model complexity, and improves training stability.

\item \textbf{Progressive voxel generation for improved generalization.} We first generate coarse-resolution voxels and then refine them to higher resolution conditioned on boundary features. By exploiting discretization-induced regularization in the coarse stage, the framework mitigates overfitting during high-resolution refinement and achieves better generalization across diverse hole boundaries.
\end{enumerate}

\section{Related Work}
\label{sec:rela}
\subsection{Traditional parameterization methods}

Traditional parameterization methods, such as nearest-plane (NP) projection~\cite{Liu12} and mean value coordinates (MVC) parameterization~\cite{floater2003}, have been widely adopted in the hole-filling and surface reconstruction literature~\cite{Varady16,Vaitkus21,kummer2025,vaitkus2025}. NP calculates a plane that minimizes the sum of distances to all points, and projects the hole boundary onto this plane to generate pcurve. MVC realizes parametric by constructing non-negative, normalized weights dependent on the angles and edge lengths from the point to each vertex and taking the weighted sum of the polygon vertices.

Despite their efficiency and simplicity, NP and MVC may fail to capture local geometric features of the hole boundary, leading to an uneven distribution of boundary samples across the parameter domain. This deficiency becomes pronounced for boundaries containing strong concavities/convexities or rapidly varying curvature, where multiple 3D regions can collapse into a narrow neighbourhood in the 2D domain under projection or coordinate-based mapping. Consequently, the trimming boundary derived from the pcurve may not faithfully match the original hole boundary, thereby degrading the accuracy and stability of subsequent constrained optimization.

To address regions of the hole that are overly concentrated in the parameter domain, the total energy function applies substantial local adjustments, disrupting the balance between the surface energy and the constraint energy. Eventually, the filled surface develops prominent folds due to the excessive accommodation of local constraints, thereby undermining the overall surface's fairness. In severe cases, it may even cause surface self-intersections or complete failure of the filling process. UVTran utilizes cross-attention to enable control points on the projection surface to focus more on adjacent hole boundary information, thereby capturing local geometric details. It generates surfaces that conform to the geometric properties of the hole boundary for projection, and then produces pcurve.

\subsection{Learning based parameterization methods}

Learning-based parameterization has gained significant traction in curve-fitting applications. For instance, Wen et al.~\cite{wen2024} use Deep Neural Networks (DNNs) to predict the knot vectors required for curve fitting, while Zou et al.~\cite{zou2025} employ transformer architectures to predict both the parameter values for fitting points and the corresponding knot vectors simultaneously. However, Zou et al. also acknowledge a limitation in their approach: extending their parameterization method to the 2D parameterization of 3D points and curves proves challenging.

To the best of our knowledge, this paper is the first to apply deep learning techniques specifically to the parameterization problem involved in filling N-sided holes with a single trimmed surface. Our approach addresses this limitation by proposing a novel method that first generates a reliable projection surface. The hole boundary is then projected onto this surface, enabling the extraction of the 2D parameter curve (pcurve) corresponding to the 3D hole boundary.

Before being fed into the model, the hole boundary is first sampled and converted into discrete points. These discrete points are analogous to inputs in point cloud learning, enabling the use of established techniques from this domain. Specifically, we draw on point cloud learning networks~\cite{qi2017, zhao2021, wu2022, chen2023, wu2024, rong2024} to extract point features.

In point cloud processing, methods such as PointNet++~\cite{qi2017++}, PointMLP~\cite{marethinking}, and PointNeXt~\cite{qian2022} are designed to handle unordered point clouds, often with semantic shapes like chairs or tables. However, in the context of hole filling, the points are ordered and correspond to abstract curves, introducing a different challenge. Therefore, in our proposed method, UVTran, we use PointNet to embed the discrete points, but modify the processing pipeline by employing traditional transformer methods for feature extraction and processing. This enables the model to generate a reliable projection surface.

After projecting the hole boundary, we obtain projected spatial curves.
The corresponding parametric representation of these projected curves
on the initial surface is used as the target pcurve, thereby effectively solving the hole-filling problem with an efficient and novel approach.

\section{Preliminaries}
\label{sec:pre}
The surface used for projection generated by the model, as well as the final filled surface, is represented using a B-spline surface. A B-spline surface is defined as follows:
\begin{equation}
S(u,v)={\sum_{i=0}^{n}\ \sum_{j=0}^{m}\ N_{i,p}(u)N_{j,q}(v)CP_{i,j}},
\label{bsp}
\end{equation}
where $N_{i,p}(u)$ denotes the degree-$p$ B-spline basis function (and similarly for $N_{i,q}(v)$) , $CP_{i,j}$ represent the control points. The objective of hole filling is to solve for the control points such that the resulting surface satisfies the geometric-continuity constraints in Eq.~\ref{geo} and preserves surface fairness.

The core idea of the proposed method is to construct a filled surface that meets industrial requirements by minimizing an energy functional. First, a surface-energy term is introduced~\cite{welch1992}. This term promotes surface fairness and discourages excessive stretching or deformation. Second, to enforce geometric continuity between the trimming boundary of the filled surface and the hole boundary, we add a constraint-energy term. A prerequisite for formulating this constraint term is the accurate computation of the pcurve on the filled surface.

The surface energy is a sum of several terms:
\begin{equation}
\begin{split}
\label{surf}
{E_{{\rm{Surf}}}} = {E_{{\rm{roc in bending}}}} + {E_{{\rm{bending}}}} ,
\\
E_{{\rm{roc in bending\;}}} =
    \\
     \int\int \left( {S_{uuu}^2 + 3S_{uuv}^2 + 3S_{uvv}^2 + S_{vvv}^2} \right)dudv,
    \\
    {E_{{\rm{bending}}}} = \int\int \left( {S_{uu}^2 + 2S_{uv}^2 + S_{vv}^2} \right)dudv,
\end{split}
\end{equation}

Here, $E_{roc in bending}$ defines a surface’s resistance to the rate 
of change in bending, $E_{{\rm{bending}}}$ represent the surface’s resistance to bending.  $S_{uu}$, $S_{uv}$, $S_{vv}$ and $S_{uuu}$, $S_{uuv}$, $S_{uvv}$, $S_{vvv}$ denote the second- and third-order partial derivatives of the surface, respectively.

The constraint energy, which enforces geometric continuity, comprises three components:
\begin{equation}
\begin{split}
  E_{Cons} = E_{Position} + E_{Normal} + E_{Curvature}, 
  \\
E_{geo} = \int {\left( B^{geo}_{trim}(t) - B^{geo}_{hole}(t) \right)^2}dt. 
\end{split}
\end{equation}
The total energy is defined as the sum of the surface and constraint energies. In the optimization, the variables are the control points of the hole-filling B-spline surface, the knot vectors, and the boundary pcurve. Welch and Witkin~\cite{welch1992} convert the total energy into matrix form for computation.
\begin{equation}
\begin{split}
\label{energy}
{E_v} = {E_{{\rm{Surf}}}} + {E_{{\rm{Cons}}}}
={\mathbf{CP}^T}(\mathbf{A}+\mathbf{D}) \mathbf{CP}-2 \mathbf{b} \mathbf{CP}+C,
\end{split}
\end{equation}
where $\mathbf{CP}$ denotes the stacked vector of control points of the filled surface. $\mathbf{A}$ corresponds to the surface-energy term, and $\mathbf{D}$ encodes the constraint contributions induced by the trimming boundary. Both $\mathbf{A}$ and $\mathbf{D}$ depend on the knot vectors and the boundary pcurve, while $\mathbf{b}$ is determined by the prescribed hole boundary geometry. The detailed derivations and computational procedures are provided in the appendix ~\ref{sec:Appendix}. Taking the partial derivative of the total energy with respect to $\mathbf{CP}$ and setting it to zero (Eq.~\ref{par}) yields the optimal $\mathbf{CP}$. This solution satisfies the aforementioned constraints. This formulation indicates that the resulting filled surface is highly sensitive to the boundary pcurve.
\begin{equation}
\label{par}
\begin{split}
    \frac{\partial E_{v}}{\partial \mathbf{P}}=2(\mathbf{A}+\mathbf{D}) \mathbf{P}-2 \mathbf{b}=0,
    \\
    \mathbf{P}=(\mathbf{A}+\mathbf{D})^{-1} \mathbf{b}.
\end{split}
\end{equation}

Similar to~\cite{Qin23,Hettin20,Karvciaus20,Karvciauskas21,Vaitkus21}, the main contribution of this paper lies in addressing the geometric challenges of N-sided hole filling. Notably, our approach focuses on ensuring geometric continuity and smoothness during the filling process. All input data consist of N-sided holes with valid topology, where each hole is defined by continuous surface patches. Specifically, boundary sampling is free from noise and irregular sampling, thereby ensuring the quality and accuracy of the input data. Consequently, the proposed algorithms and models operate on geometric models that are topologically sound and free of geometric defects.

\begin{figure*}[t]
  \centering
\includegraphics[width=0.9\textwidth]{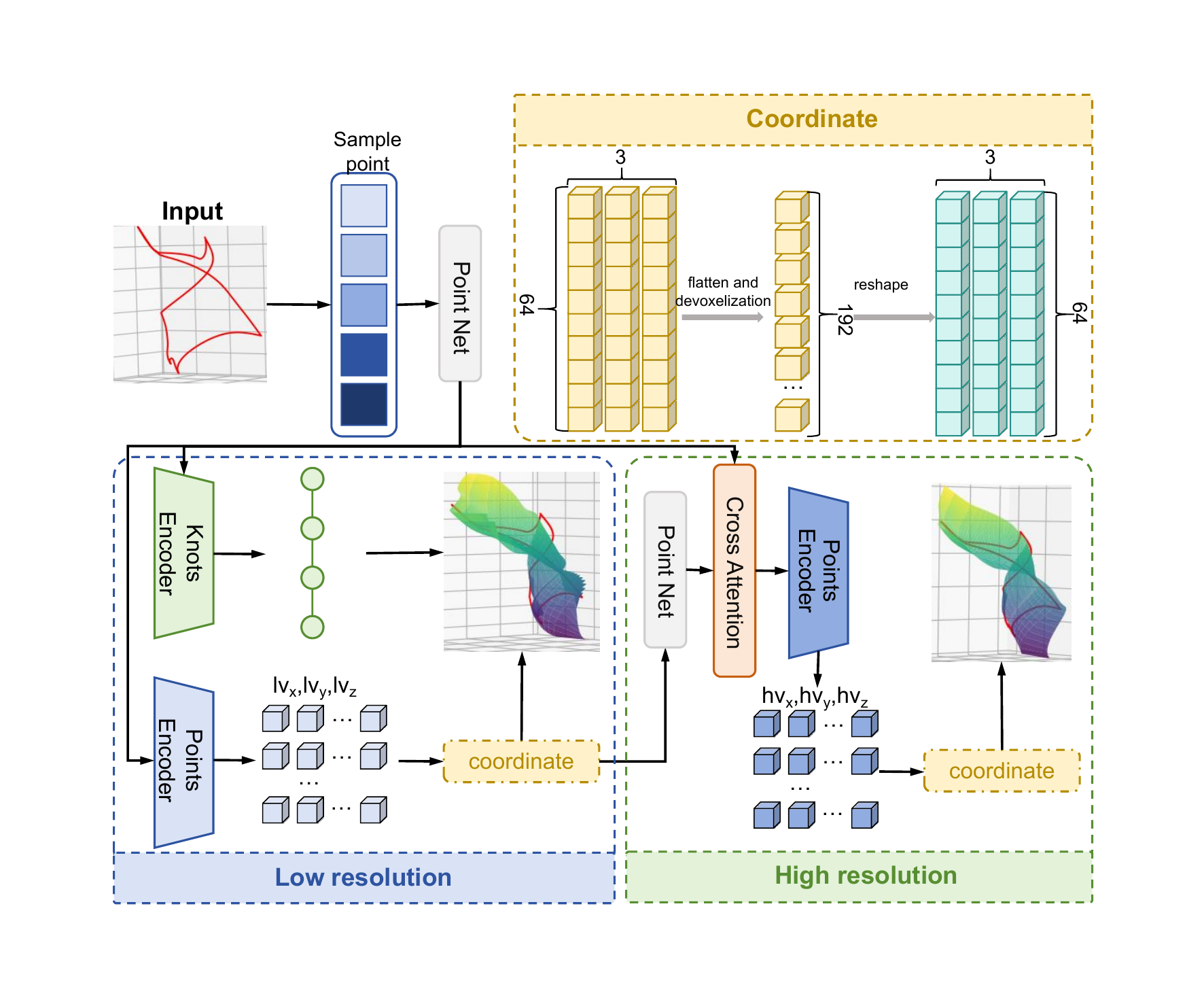}
\vspace{-10mm}
\caption{\textbf{UVTran architecture.} Boundary samples are embedded by PointNet and aggregated by a transformer; two decoding heads regress knot vectors and control points, which are coordinated to form the predicted projection surface.}
\label{fig:net}
\end{figure*}

\section{Method}
\label{sec:method}

\begin{algorithm}

\caption{Hole-Filling Algorithm}\label{algorithm:hf}
\begin{algorithmic}[1]
\State \textbf{Input:} Hole Boundary geometry $B^{\mathrm{geo}}_{\mathrm{trim}}$ , Sample Point  $P$
\State \textbf{Output:} Filling Surface $S_{fill}$
\State \textbf{Generate Projection Surface} \[S_{pro} = model(P)\] 
\State \textbf{Generate pcurve by Projecting} 
\[pcurve = S_{pro}^{-1}(\Pi_{S_{pro}}(B^{\mathrm{pos}}_{\mathrm{trim}}(t)))\]
\State \textbf{Construct Energy Function:} \[E_v = E\begin{pmatrix} pcurve,& B^{\mathrm{geo}}_{\mathrm{trim}},  & S_{fill}\end{pmatrix} \]
\State \textbf{Solve Filling Surface:}
\[
S_{fill}^* = \arg \min_{S_{fill}} E_v(S_{fill})
\]

\end{algorithmic}
\end{algorithm}

\paragraph{Notation}
$P$ denotes the coordinates of sampled points on the hole boundary; $LCP$ denotes the control points predicted at the low-resolution stage, 
$HCP$ denotes the control points predicted at the high-resolution stage, and $CP$ denotes the control points of the filling surface. $S_{pro}$
denotes the projection surface generated by the model; $S_{fill}$ denotes the final generated filling surface. $\Pi$ denotes the projection operator; $S^{-1}$
denotes the inverse mapping for parameter recovery;
$model(\cdot)$ denotes the deep-learning inference process; 
$F$ denotes the corresponding features extracted by the network; and $V_i$ denotes the voxel value; $\Delta v$ denotes the voxel resolution, $lv$ denotes the voxels predicted by the low-resolution stage model, and $hv$ denotes those predicted by the high-resolution stage model.

This section first provides an overview of the UVTran model architecture and its core design principles. It then describes the implementation of voxelization, cross-attention (CA), and the progressive-resolution (PR) training strategy, and finally defines the loss functions used for model training.

Fig.~\ref{fig:net} illustrates the UVTran architecture. Voxelization is the inverse process of coordinate. We first sample the hole boundary to obtain an ordered point set $P\in\mathbb{R}^{n\times 3}$. Although each point can be associated with differential attributes such as normals and second-order terms, we use coordinates as the default input. This choice is motivated by (i) the practical availability and robustness of coordinates in industrial pipelines, and (ii) the fact that under noise-free sampling from smooth surfaces, these differential attributes are strongly correlated with the local coordinate configuration and can be implicitly captured by neighborhood-based feature aggregation. We further validate this design choice in the ablation study (Sec.~\ref{subsection:abstudy}). The sampled points are embedded using a lightweight PointNet to produce boundary features $F_p\in\mathbb{R}^{n\times d}$. The features $F_p$ are then processed by two parallel $l$-layer encoders: a control-point encoder and a knot-vector encoder. Their outputs are fed into a multilayer perceptron (MLP) to predict an initial set of control points (represented in voxelized form; see Sec.~\ref{subsec:voxel}) and the corresponding knot vectors.

To refine the initial control points $LCP$, we embed the predicted initial control point coordinates using the same lightweight PointNet to obtain $F_{lcp}$. Next, $F_{lcp}$ and the boundary features $F_p$ are fed into a cross-attention (CA) module, which learns local correspondences between boundary samples and candidate control points and computes attention weights that emphasize the boundary regions most relevant to each control point. The CA output is concatenated with $F_{lcp}$ along the feature dimension to form fused features. These fused features are passed through another $l$-layer encoder for deeper feature extraction, after which a second MLP outputs refined, high-precision control points (again represented in voxelized form and subsequently decoded into continuous coordinates).

Given the refined control points and knot vectors, we construct a projection surface $S_{pro}(u,v)$ that serves as an intermediate, geometrically reliable surface for parameterization. We then project the 3D hole boundary onto $S_{pro}$ to obtain the corresponding 2D pcurve in the $(u,v)$ domain. 

Then, the pcurve predicted by the model defines the trimming boundary in parameter space and, together with the 3D hole boundary geometric information (e.g., positions, normals, curvatures), is used to construct the total energy in Eq.~\ref{energy}. This energy typically includes (i) boundary-consistency terms that enforce geometric continuity with the input boundary (e.g., positional $G^0$, normal $G^1$, and curvature-related terms where applicable) and (ii) interior fairness/regularization terms that suppress oscillations and promote a smooth, physically plausible surface. Finally, we solve for the surface parameters (e.g., the control points of the final filling surface) that minimize the total energy, yielding the final filled surface.

The concrete steps of the algorithm are delineated in \textbf{Algorithm~\ref{algorithm:hf}}. In summary, UVTran predicts a projection surface and, via projection, generates a pcurve; under the guidance of the projection-generated pcurve and hole boundary geometry, an energy-based optimization is performed to obtain the final filling surface.

\subsection{Voxelization}
\label{subsec:voxel}
We discretize the 3D coordinate space by uniformly partitioning each axis into intervals of length $\Delta v$ and assigning an integer index to each interval. For a control point $CP(x,y,z)\in\mathbb{R}^3$, its voxel index along each axis is obtained as $v_i=\left\lfloor \frac{i}{\Delta v}\right\rfloor$, where $i\in{x,y,z}$, $\Delta v$ is the voxel resolution, and $\lfloor\cdot\rfloor$ denotes the floor operator. Equivalently, letting $V_N=1/\Delta v$, we obtain $v_i=\left\lfloor i,V_N\right\rfloor$.

This voxelization is performed independently along each coordinate axis, producing a 3-tuple label $(v_x,v_y,v_z)\in\mathbb{Z} \cap [0, V_N)^3$. This differs from direct 3D voxelization, which flattens the 3D grid into a single categorical index. If we instead map $(v_x,v_y,v_z)$ to a single linear index (row-major order) on a cubic grid, a typical flattening rule is $v=v_x V_N^2+v_y V_N+v_z$, which corresponds to an $(V_N)^3$-way classification problem.

During dataset construction, we convert each ground-truth control point into its discrete voxel indices $(v_x,v_y,v_z)$, thereby casting learning as a classification task rather than direct coordinate fitting. Compared with direct 3D voxel classification, per-axis discretization is substantially more practical: it replaces a single $(V_N)^3$ class problem with three $V_N$ class problems. For example, when $\Delta v=0.01$, then $V_N=100$. Direct 3D voxel classification would require $100^3=1{,}000{,}000$ categories, which is computationally expensive and data-inefficient, whereas per-axis classification requires only $ 100$ classes.

Given predicted voxel indices $(v_x,v_y,v_z)$, we reconstruct a continuous point by placing it at the center of the corresponding voxel cell: 
\begin{equation}
\label{recon}
    i=(v_i+0.5)\Delta v \quad i\in{x,y,z}.
\end{equation} Centering reduces bias compared with using the interval’s lower boundary. The maximum discretization error occurs when the true point lies at a voxel corner relative to the cell center. Since each coordinate error is at most $\Delta v/2$, the Euclidean discretization error satisfies $\varepsilon_{\mathrm{dis}}\le \sqrt{3},\Delta v/2$.

In direct fitting, the network outputs continuous coordinates $(x,y,z)$ for each control point. Any deviation from the ground truth directly produces geometric error, and small perturbations in the input or intermediate features can propagate into the final coordinates. Moreover, a fitting model must learn both the point precise numeric coordinates, which can be sensitive to noise, outliers, and scale variations across datasets.

In contrast, the voxel-based formulation predicts discrete labels (per-axis indices) via classification. The network outputs a probability distribution over candidate indices; inference selects the most likely index (e.g., via $\arg\max$). This yields a practical advantage:

Robustness to perturbations via decision boundaries. Classification is primarily determined by the relative class probabilities. Moderate shifts in probability mass do not change the predicted index provided that the top-1 class remains unchanged. This behavior creates a natural buffer against small perturbations in the features and minor annotation inconsistencies.

Overall, per-axis voxel prediction reduces sensitivity to minor fluctuations, avoids an intractable $(V_N)^3$-class output space, and provides an explicit accuracy–complexity trade-off via $\Delta v$.  Discretization is only performed on the control point coordinates of the target surface, where the continuous coordinates are discretized into voxel coordinates to facilitate the generation of control points for the model. In this paper, the input hole boundaries, the generated projection surfaces, and filling surfaces are all represented in a continuous form.

\subsection{Low-resolution Stage}
\label{subsec:lowre}

While a smaller $\Delta v$ (e.g., $0.01$) can reduce discretization error, using such a fine resolution too early increases prediction variance and the risk of voxel-label errors. Therefore, we adopt $\Delta v=0.1$ in the low-resolution stage to improve the stability and reliability of the initial solution, and defer fine-grained accuracy to subsequent refinement stages.

We first sample points along the boundary of the N-sided hole, denoted by $\partial\Omega\subset\mathbb{R}^3$, to obtain an ordered (i.e., consistently indexed) point set $P=[p_1,p_2,\ldots,p_n]\in\mathbb{R}^{n\times 3}$. A lightweight PointNet-style encoder $\Phi(\cdot)$ is then used to embed these boundary samples into point-wise features, $F_p=\Phi(P)$ with $F_p\in\mathbb{R}^{n\times d}$, where $\Phi$ is implemented as a three-layer MLP with shared weights across points and $d$ denotes the feature dimension.

The target surface at the low-resolution stage is represented by a $8\times 8$ control grid (i.e., 64 control points). And all target surfaces are cubic clamped B-spline surfaces, so the knot vectors in both the U and V directions have a length of 12. The first and last knots are each repeated four times, with the initial knot set to 0 and the final knot set to 1. As a result, only the four intermediate knots need to be predicted. We map the point-wise boundary features to a set of 64 control point tokens and 4 knot vector tokens via linear projection.
\begin{equation}
    \begin{split}
        F_{cp}=\mathrm{Linear}(F_p) \quad F_{cp}\in\mathbb{R}^{64\times d},\\
        F_{knot}=\mathrm{Linear}(F_p) \quad F_{knot}\in\mathbb{R}^{4\times d}.
    \end{split}
\end{equation}

These tokens are processed by two parallel $l$-layer Transformer encoders: Point encoder $\Psi_p$ and Knot encoder $\Psi_k$. All these encoders are basic Transformer encoders with self-attention mechanisms. Both encoders use self-attention to model global correlations among the control points and knot vectors tokens, producing two representations: 
\begin{equation}
    \begin{split}
        F_{ptran}=\Psi_p(F_{cp}) \quad F_{ptran}\in\mathbb{R}^{64\times d},
        \\F_{ktran}=\Psi_k(F_{knot})\quad F_{ktran}\in\mathbb{R}^{4\times d}.
    \end{split}
\end{equation}

Finally, two MLP heads predict voxelized control point coordinates and knot vectors: 
\begin{equation}
    \begin{split}
        \mathrm{VLogits}=v\mathrm{MLP}(F_{ptran}) \quad \mathrm{VLogits} \in\mathbb{R}^{64\times3V_N }, \\ \mathrm{Knot}=k\mathrm{MLP}(F_{ktran})\quad \mathrm{Knot}\in\mathbb{R}^{4}.
    \end{split}
\end{equation}

Following the per-axis voxelization strategy, each control point predicts a discrete voxel index for $x$, $y$, and $z$ separately. Concretely, the voxel head outputs logits that are reshaped into $[192,V_N]$, and a softmax is applied along the voxel dimension, yielding $V_p=\mathrm{Softmax}(\mathrm{VLogits})$, where $V_N=1/\Delta v$ is the number of bins per axis at voxel resolution $\Delta v$. For each control point and each coordinate axis, we select the most probable voxel via $ argmax $, yielding voxel indices $(v_x,v_y,v_z)$. These indices are then mapped back to continuous coordinates using the voxel center reconstruction rule (Eq.~\ref{recon}), producing the initial control points $LCP\in\mathbb{R}^{64\times 3}$.

The goal of the initial stage is not to recover fine geometric detail immediately, but to generate a stable and globally plausible control mesh that serves as a reliable starting point for subsequent refinement. Using a coarser voxel resolution (here $\Delta v=0.1$) supports this goal in several ways:

Reduced classification difficulty and higher top-1 reliability. With axis-wise voxel classification, each coordinate corresponds to a $V_N$ way classification problem. When $\Delta v=0.1$, $V_N=10$, whereas when $\Delta v=0.01$, $V_N=100$. Increasing from 10 to 100 classes per axis substantially enlarges the decision space and increases the risk of misclassification, particularly early in training when features are less discriminative. A coarse grid therefore increases the likelihood of selecting the correct voxel bin, which is critical for producing usable initial control points.

Prioritizing global structure before local detail. In hole filling, the first requirement is typically global shape consistency—e.g., producing control points that roughly span the hole with the correct overall orientation and scale. Fine-scale adjustments (e.g., local curvature and small boundary concavities) are better handled in subsequent stages with higher resolution. A coarse resolution encourages the network to learn these global cues first.

More efficient training and inference. Fewer classes per axis reduce memory and computation in the final prediction layers (logits, softmax, and loss), enabling faster convergence at the coarse stage and allowing model capacity to focus on learning meaningful global correlations in the Transformer rather than handling an overly granular discretization early in training.

While a smaller $\Delta v$ (e.g., $0.01$) can reduce discretization error, using such a fine resolution too early increases prediction variance and the risk of voxel-label errors. Therefore, we adopt $\Delta v=0.1$ in the low-resolution stage to improve the stability and reliability of the initial solution, and defer fine-grained accuracy to subsequent refinement stages.

\subsection{Cross Attention}
\label{subsec:ca}
B-spline surfaces are characterized by a notable locality property, where each control point influences the surface only within its neighboring region, without affecting the global properties of the entire surface~\cite{piegl2012}. This property allows for fine adjustments to surface morphology, making B-splines particularly well-suited for modeling complex geometric features with localized control.

In this approach, low-resolution control points $LCP$ are embedded using the same PointNet architecture applied to the boundary sampling points, as detailed in Section~\ref{subsec:lowre}. The features of the control points serve as queries, while the hole boundary point features $F_p$ are used as both keys and values in the cross-attention mechanism. The resulting weighted features are computed by processing the input features through the attention mechanism, as described by the following equation:
\begin{equation}
    \begin{split}
    \hat{Q} = F_{lcp} \quad
    \hat{K} =\hat{V} = F_{p},
    \\
F_{w}=\operatorname{Softmax}(S)=\operatorname{Softmax}\left(\frac{\hat{Q} \cdot \hat{K}^{T}}{\sqrt{d_{\text {model }}}}\right)\hat{V}.
    \end{split}
\end{equation}
Cross attention effectively captures the localized relationships between control points and the boundary, allowing for precise feature weighting. This locality-preserving property of cross-attention enhances the model's ability to adjust the surface features without disrupting the global geometry, providing a more efficient and accurate approach to modeling complex shapes.

\subsection{High-resolution Stage}
\label{subsec:highre}
To fully leverage the information from the initial control points, the weighted features are concatenated with the original control point features, resulting in a fused feature vector $F_{\text{fuse}}$:
\begin{equation}
        F_{fuse} = concate(F_{w},F_{lcp}) \in R^{64,dim*2}.
\end{equation}

This fused feature $F_{\text{fuse}}$ is then processed through an additional set of $l$-layer Transformer encoders. This process can be viewed as a refinement and optimization of the initial solution, constrained by the boundary conditions~\cite{yang2018}. After passing through a multi-layer perceptron (MLP), high-resolution control point voxels $H_{V_p} \in \mathbb{R}^{192 \times 10 \times V_N}$ are generated.

As with the initial control points, voxel selection is performed probabilistically. The selected voxels are then mapped to 3D coordinates $HCP \in \mathbb{R}^{64 \times 3}$ for further processing. During the high-resolution phase, the model predicts only the voxelized representation of the control points. The final projection surface is constructed using the initially generated knot vectors.

In the low-resolution training phase, discrete errors $\varepsilon_{\text{dis}}$ persist even after the model has fully converged. These errors enhance the model’s robustness to potential distribution shifts in the test set during the high-resolution phase, thus improving generalization performance. If training were instead initialized with high-resolution data, the discrete errors would be minimal at convergence. However, this would lead to a model highly sensitive to distribution shifts, resulting in higher generalization errors. By utilizing the discrete errors $\varepsilon_{\text{dis}}$ as an optimization target during the high-resolution phase, the model’s ability to generalize is enhanced, leading to more robust performance on unseen data.

\subsection{Loss Function}
The generation of the knot vector employs the mean squared error (MSE), while the generation of control points voxel utilizes the cross-entropy loss function. The specific formulations are detailed as follows. The overall loss function is the sum of these individual loss functions.
\begin{equation}
\begin{split}
        L_{MSE}=\frac{1}{m} \sum_{i=1}^{m}\left(\hat{k}_{i}-k_{i}\right)^{2},\\
    L_{CE}=-\frac{1}{n} \sum_{i=1}^{n} \sum_{c=1}^{C} v_{i, c} \log \left(\hat{v}_{i, c}\right),
    \\
    L = L_{MSE}+L_{CE}.
\end{split}
\end{equation}
Where $k_{i}$ denotes the target knot vector, $\hat{k}_{i}$ the predicted knot vector, $v_{i, c}$ the voxelized coordinates of the target control points, and $\hat{v}_{i, c}$ the voxelized coordinates of the predicted control points. $m$ denotes the number of knot vectors, $n$ denotes the number of control points, and $C$ denotes the number of voxel. In the high-resolution stage, only the cross-entropy loss function is adopted.

\section{Experiments}

\begin{figure}[]
  \centering
\includegraphics[width=0.5\textwidth]{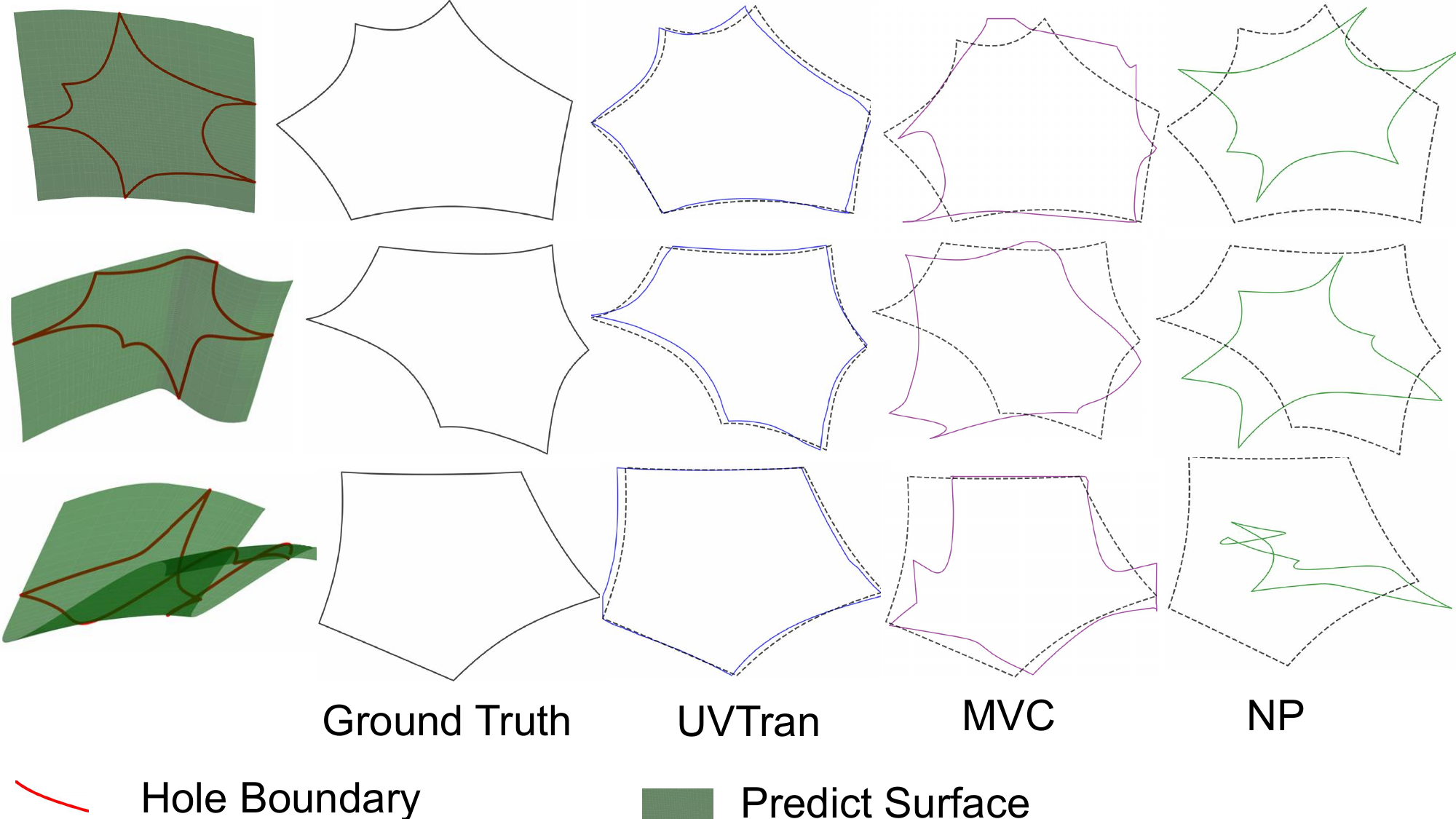}
     \caption{\textbf{3D curve parameterization.} Comparison of different methods on the test set.}

\label{fig:param}
\end{figure}

\begin{figure*}[]
  \centering
\includegraphics[width=0.9\textwidth]{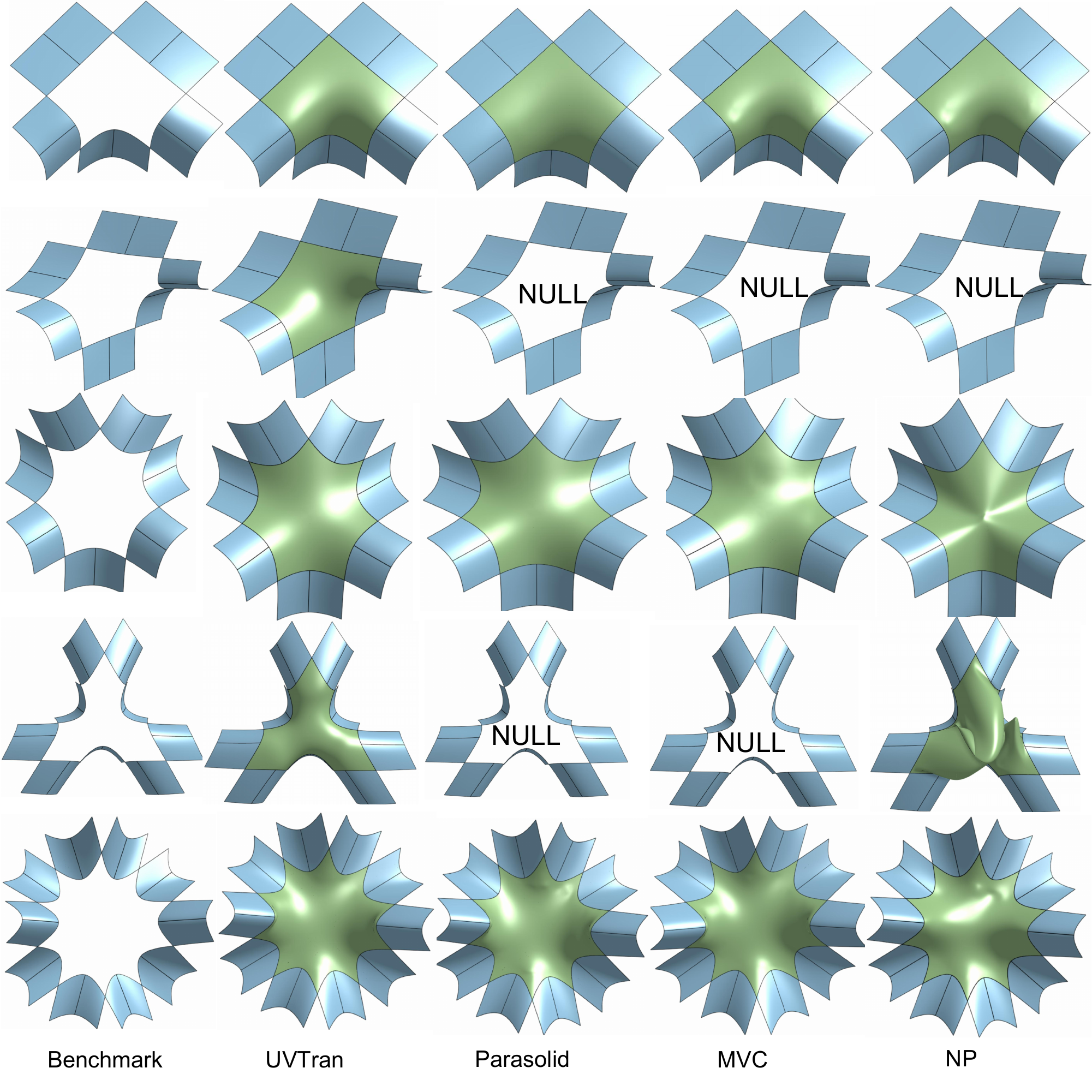}
     \caption{\textbf{Hole-filling comparison.} Qualitative results of different methods on the benchmarks.}

\label{fig:com}
\end{figure*}

\begin{figure*}[]
  \centering
\includegraphics[width=0.9\textwidth]{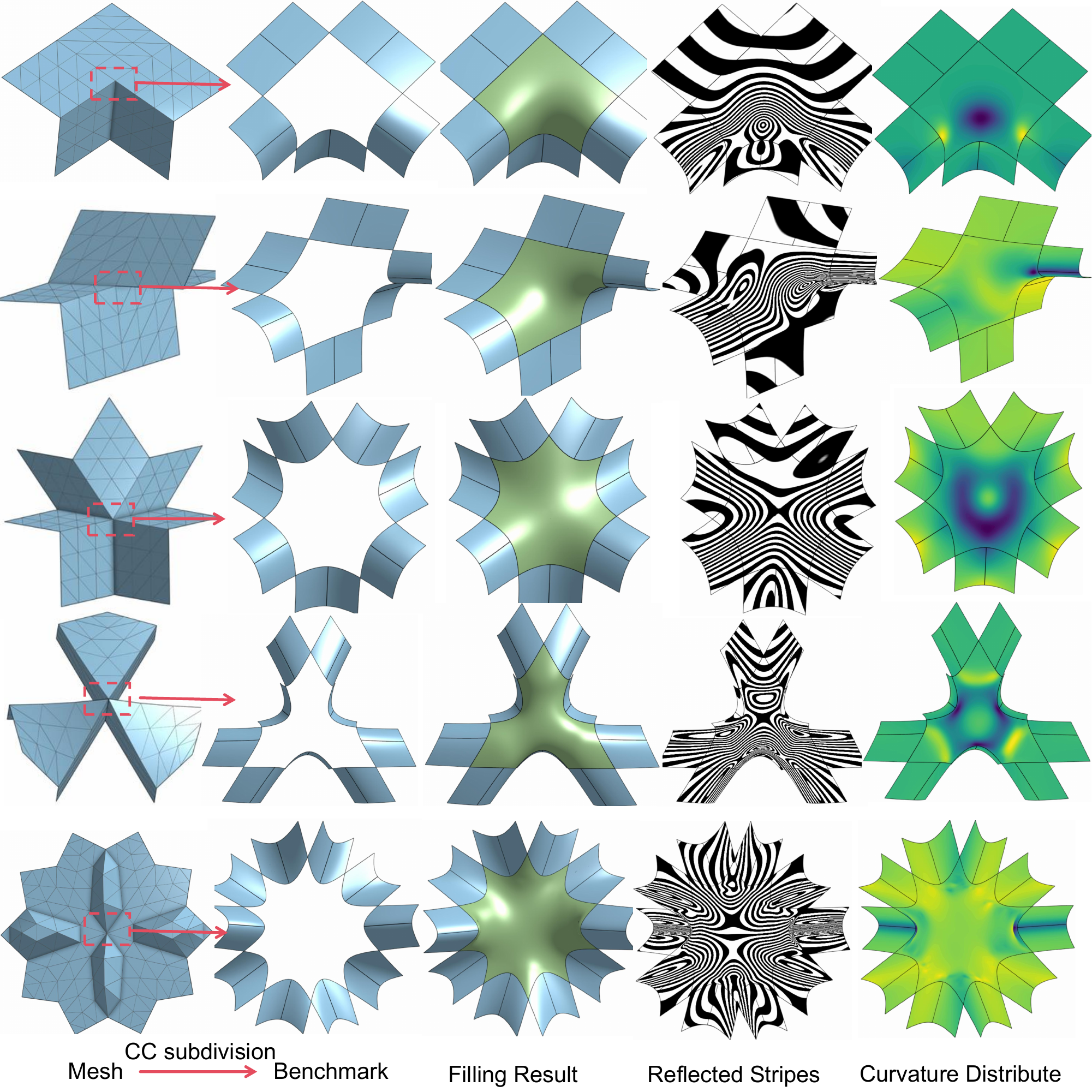}
    \caption{\textbf{Benchmark visualization.} From left to right: initial mesh, hole-boundary benchmark (produced via a subdivision mesh), filling result, reflection stripes, and curvature distribution.}
\label{fig:result}
\end{figure*}

\begin{figure}[]
  \centering
\includegraphics[width=0.5\textwidth]{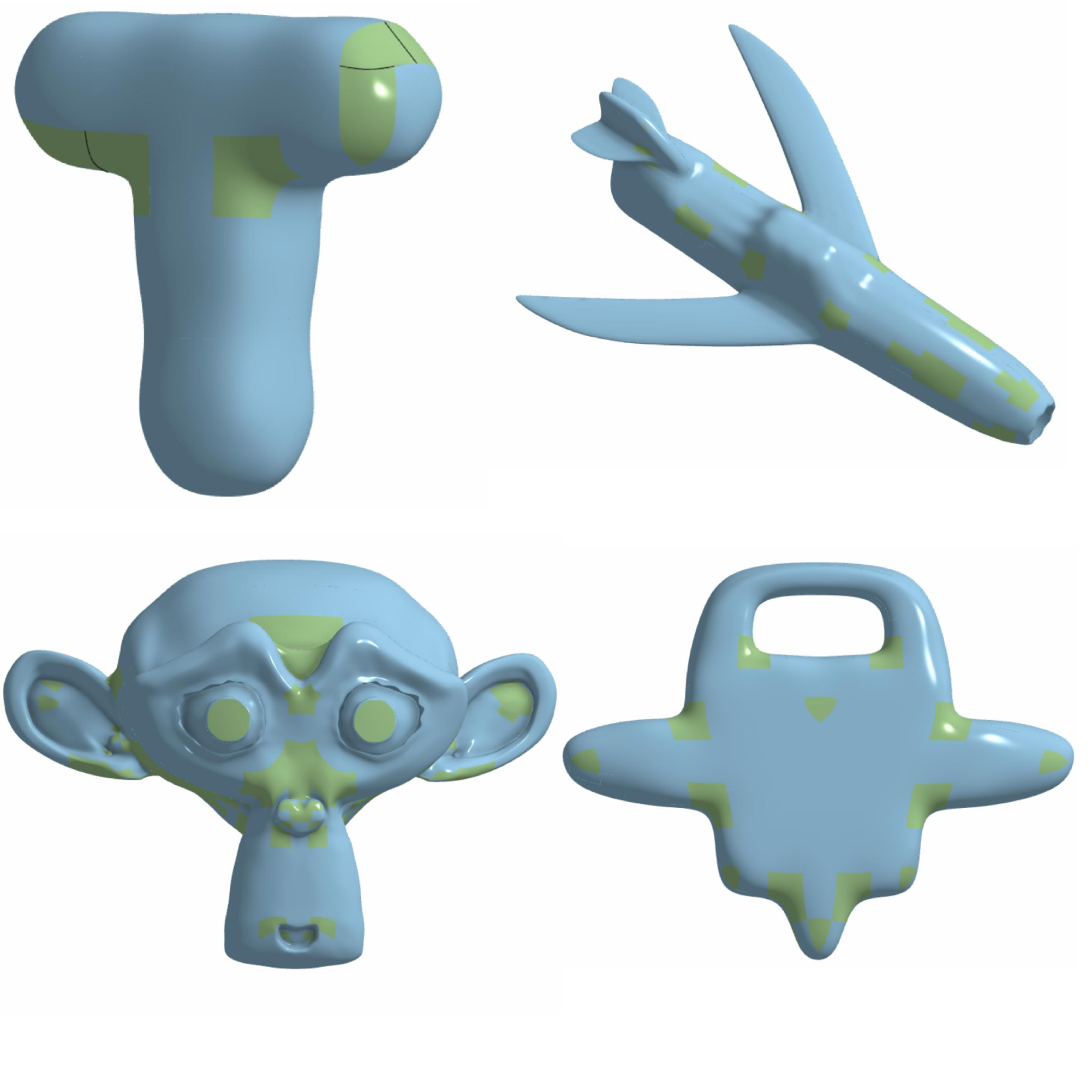}
\vspace{-10mm}
    \caption{\textbf{Real-world models.} Real-world geometric models with the filled surfaces highlighted in green.}
\label{fig:bigmodel}
\end{figure}

\begin{figure}[]
  \centering
\includegraphics[width=0.5\textwidth]{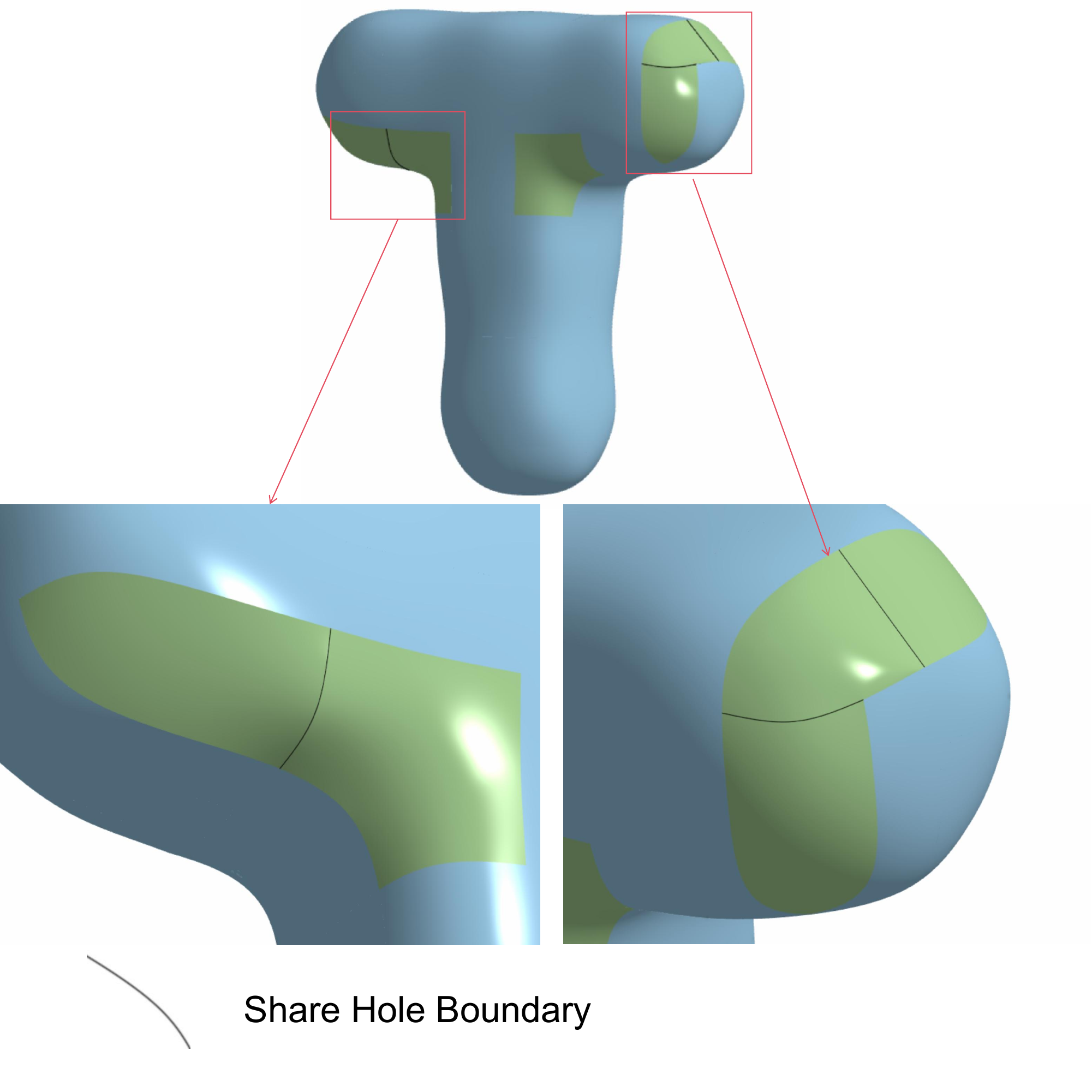}
    \vspace{-8mm}
    \caption{\textbf{Multiple adjacent holes.} Hole-filling results for multiple adjacent holes; black curves denote shared boundaries between holes.}
\label{fig:multi}
\end{figure}

\subsection{Datasets}
Our dataset consists of two parts. The first part is collected from ~\cite{3Dscanrep} and is used to construct the training and test sets, while the second part is obtained from ~\cite{surflab} and is used to build the benchmark. The benchmark is constructed from out-of-distribution data with respect to the training set. The details are described as follows.

\begin{figure}[]
  \centering
\includegraphics[width=0.5\textwidth]{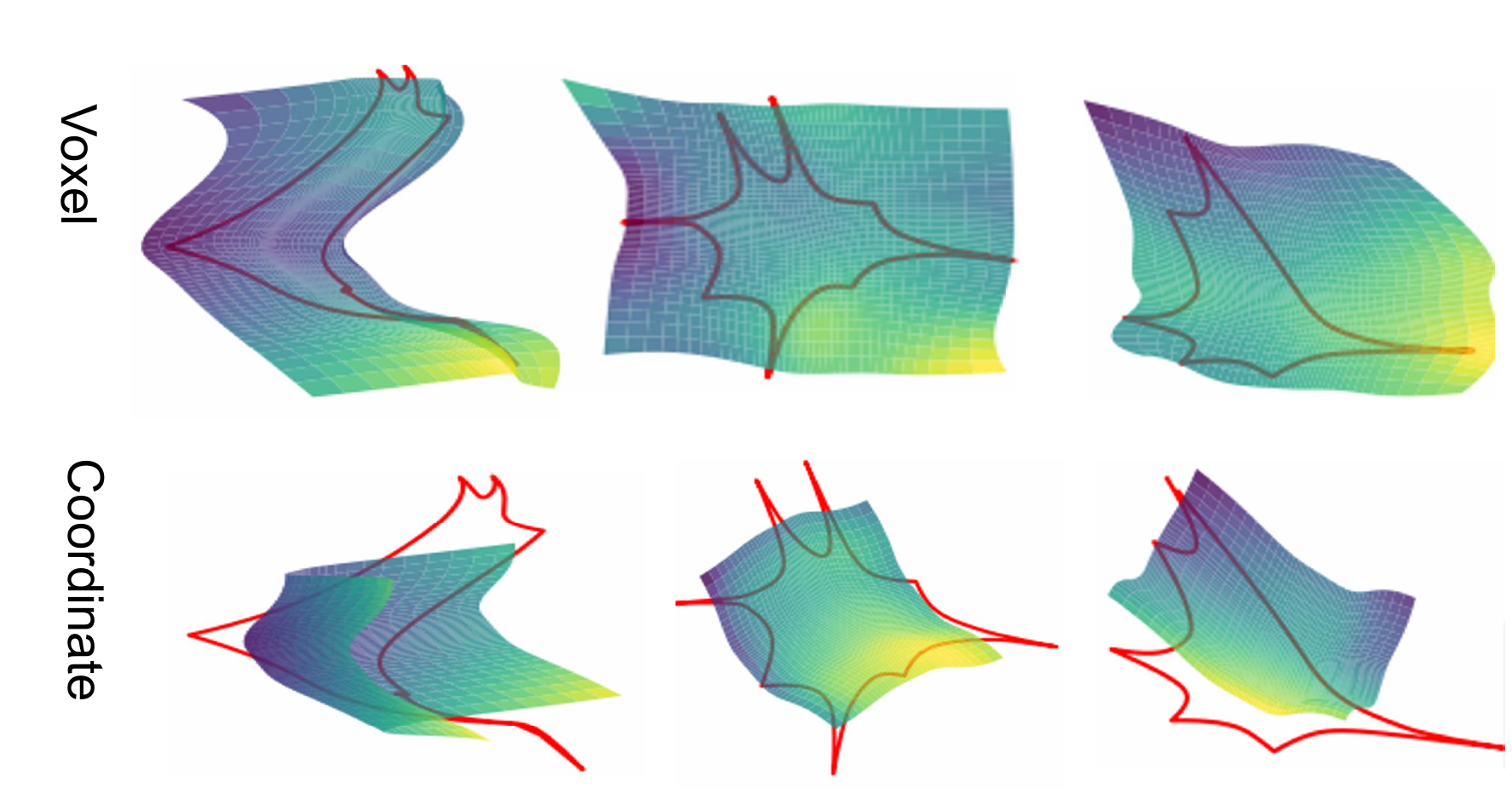}
     \caption{\textbf{Voxel and coordinate representations.} Voxelization result and coordinate-based representation; the red curve indicates the hole boundary.}
\label{fig:vor}
\end{figure}

\textbf{Train and Test Datasets.}
The model data in the Stanford 3D Scanning Repository~\cite{3Dscanrep} consists of meshes. To construct the training dataset, the Catmull-Clark Subdivision~\cite{Catmull98} algorithm is first applied to transform the meshes into multiple B-spline surfaces. For each surface patch, an 8×8 grid of points is sampled to serve as control points, and random knot vectors are generated concurrently. By following these steps, a total of 30,000 initial surfaces are generated.

These 30,000 surfaces selected from the Stanford 3D Scanning Repository~\cite{3Dscanrep} have undergone normalization. Additionally, 10000 closed pcurves corresponding to real hole boundaries have been included. The core task is to generate projection surfaces, as the shape of the surface significantly impacts the experimental results. While the influence of the pcurves is relatively limited, there is no need to select an excessive number of them. In the dataset construction phase, 10 pcurves from the aforementioned set are randomly selected for each surface and used to trim it. The generated trimming curves are employed to simulate the hole boundary, and the voxelized control points and knot vectors of the trimmed surfaces are taken as target values. The surfaces and pcurves form the optimal solution under the current boundary conditions, ensuring zero error with respect to geometric continuity.

Before each trimming, the surface undergoes noise addition while maintaining fairness. Adding noise increases the diversity of surfaces, enhances the model's generalization ability, reduces the impact of data fluctuations, and improves robustness. The target surface selected in this study is a third-order B-spline surface with $8\times 8$ control points, which is capable of adequately representing the geometric features of complex models. Knot vectors in both the U and V directions have a length of
12. The first and last knots are each repeated four times, with
the initial knot set to 0 and the final knot set to 1. Experimental results demonstrate that the projection surface with this configuration effectively handles complex hole boundaries, proving its effectiveness and robustness.

90\% of the surfaces and curves were used to construct the training set, while the remaining 10\% were used to build the test set.  In the training and testing processes, the hole boundary surfaces, generated projection surfaces, and filling surfaces are all represented in a continuous form. Grid data must be converted into continuous data via a subdivision algorithm before being used as the input to the algorithm.

\textbf{Benchmark.}
In Fig.~\ref{fig:result}, we present a variety of geometric structures with challenging characteristics, where the generated trimmed filling surfaces are surrounded by a ring of regular bicubic patches. These structures are derived from SurfLab~\cite{surflab}, a comprehensive dataset widely used in the field of geometric modeling. Each geometric structure is carefully constructed by applying a single Catmull-Clark subdivision step, resulting in an N-sided hole that serves as a typical test case for hole-filling algorithms.

Many prominent hole-filling methods, such as those proposed by Qin et al.~\cite{Qin23}, Hettin et al.~\cite{Hettin20}, Karvciauskas et al.~\cite{Karvciaus20}, Karvciauskas et al.~\cite{Karvciauskas21}, and Vaitkus et al.~\cite{Vaitkus21}, have used these datasets as benchmark evaluations, demonstrating their utility for comparing different approaches. Additionally, we extend the evaluation to geometric models with multiple holes simultaneously, as well as to real-world models, which often exhibit more complex boundary conditions and geometric features. This expanded testing ensures that our method can handle a broad range of practical and theoretical scenarios.

\begin{table}[t]
\centering
\caption{\textbf{Two-stage model efficiency.} Performance and efficiency of the two-stage models.}
\label{tab:efficiency}
\begin{tabular}{lccc}
\toprule
Stage & Parameters (M) & FLOPs (G) & FPS \\
\midrule
Low-resolution  & 6.44 & 0.44 & 275.84 \\
High-resolution & 7.01 & 0.45 & 257.50 \\
\bottomrule
\end{tabular}
\end{table}
The results of our tests are shown in Fig~\ref{fig:bigmodel}, where the green regions highlight the filled surfaces generated by our proposed method. For multiple connected holes, after providing the geometric information for each boundary of every hole, each hole is processed individually, as shown in Fig~\ref{fig:multi}. 

\begin{table}[t]
\centering
\caption{\textbf{Parameter error comparison.} Parameter error on the test dataset compared with state-of-the-art methods.}
\label{tab:pe}
\begin{tabular}{cc}
\toprule
Method & Parameter error \\
\midrule
UVTran & \textbf{5.84e-3} \\
MVC    & 1.07e-2 \\
NP     & 9.25e-2 \\
\bottomrule
\end{tabular}
\end{table}

\begin{table}[t]
\centering
\caption{\textbf{STR comparison.} STR (\%) on the benchmark set compared with state-of-the-art methods.}
\label{tab:tol}
\begin{tabular}{cccccc}
\toprule
   & UVTran & Parasolid & MVC & NP \\
\midrule
  $G^0$ & \textbf{100} & \textbf{100} & 86 & 81 \\
  $G^1$ & \textbf{95}  & 83           & 72 & 67 \\
  $G^2$ & \textbf{90}  & 80           & 68 & 61 \\
\bottomrule
\end{tabular}
\end{table}

\begin{table}[htbp]
\centering
\caption{\textbf{Voxel resolution sensitivity.} Parameter error under different voxel resolutions on the test dataset.}
\label{tab:pr}
\begin{tabular}{cc}
\toprule
Resolution & Parameter error \\
\midrule
1e-1, 1e-1 & 3.93e-2 \\
1e-1, 1e-2 & \textbf{5.84e-3} \\
1e-1, 1e-3 & 7.55e-3 \\
1e-2, 1e-2 & 2.87e-2 \\
1e-2, 1e-3 & 1.68e-2 \\
1e-3, 1e-3 & 8.11e-2 \\
\bottomrule
\end{tabular}
\end{table}

\subsection{Implementation details}
During the experiment, both training and inference were conducted on a single NVIDIA RTX 5090 graphics card. The dropout coefficient was set to 0.05. The positional, normal, and curvature tolerances are set to 1e-6, 1e-3, and 1e-1. Additional implementation details can be found in our open-source code.

Our two-stage models are designed with a focus on efficiency. Parameters, FLOPs, and FPS are summarized in ~\ref{tab:efficiency}. In terms of performance, the models deliver 275.84 and 257.50 frames per second (FPS), respectively. These results demonstrate that the models strike a balance between lightweight design and high performance, making them well-suited for real-time applications where efficiency is critical.

\subsection{Evalution Metics}
\textbf{Parameter error.} In the test dataset, parameter error is computed by the mean L2 norm of the discrepancy between the parameters generated by different parameterization methods and the ground truth.

\textbf{Satisfy tolerance rate.} The Satisfy Tolerance Rate (STR) refers to the proportion of cases where the geometric deviation between the filled surface and the hole boundary meets the preset engineering tolerance threshold. This metric reflects the engineering feasibility of the filling results.

\textbf{Surface quality. }
Due to the lack of a universally accepted standard for surface quality assessment, this study adopts a qualitative evaluation approach similar to other methods ~\cite{Qin23,Hettin20,Karvciaus20,Karvciauskas21,Vaitkus21}. The primary evaluation criteria include surface continuity and smoothness. Specifically, through rendering results, we assess the geometric form of the surface from different viewpoints, ensuring that the surface is smooth and free from significant wrinkles or sharp edges. Reflection stripe analysis is employed to detect surface smoothness and geometric continuity, as the flatness and uniformity of the reflection stripes can effectively indicate whether the surface exhibits irregular fluctuations or defects. Additionally, curvature distribution maps are used to analyze the curvature variation across different regions of the surface, which helps determine whether the surface transitions smoothly to the hole boundary and ensures that excessive compression or stretching does not occur in high-curvature areas.

\subsection{Comparison with State-of-the-art Methods}
We conducted comparisons with state-of-the-art (SOTA) methods, including MVC~\cite{floater2003}, NP~\cite{Liu12}, and Parasolid~\cite{parasolid}, a commercial modeling software widely used in industry. 
Parasolid, as one of the leading geometric kernels in the industry, is widely utilized across various fields. Many renowned industrial design software packages, such as Siemens NX and SolidWorks, rely on Parasolid as the core engine for geometric processing. By comparing our results with Parasolid, we are able to ensure the authority and reliability of the experimental outcomes. Given that Parasolid is proprietary software, its parameterization process cannot be directly analyzed or modified. During the testing phase, we used identical input data and tolerance standards for Parasolid, other comparative algorithms, and the algorithm proposed in this study. To ensure a fair comparison of the different methods, we conducted tests under the same conditions for each algorithm, minimizing the impact of external factors. Specifically, when Parasolid successfully completes filling, it returns accurate results; however, if Parasolid fails to perform the filling, it returns an error code. As shown in Table~\ref{tab:pe} and Figure~\ref{fig:param}, compared to MVC, UVTran reduces the parameterization error from 1e-2 to 5e-3, demonstrating a significant improvement in accuracy.

Additionally, we performed comparative analyses of the final hole-filling results under the $G^0$, $G^1$, and $G^2$ continuity constraints to verify UVTran’s ability to maintain superior performance across different continuity levels. As shown in Table~\ref{tab:tol}, compared to Parasolid, UVTran improved the Surface Tension Ratio (STR) by 10\% under $G^2$ continuity.

Figure~\ref{fig:com} illustrates the results on the benchmark, where 'NULL' indicates failure. For challenging hole boundaries, UVTran demonstrates greater robustness and better surface quality. Figure~\ref{fig:result} showcases the UVTran-based method under $G^2$ continuity, including the rendered results, reflection stripes, and curvature distribution. The generated filling surface is fair, with a uniform curvature distribution and no abrupt changes, highlighting the superior surface quality maintained by UVTran.

\subsection{Ablation study}
\label{subsection:abstudy}
\textbf{Ablation on input modalities.} We evaluate whether explicitly providing normals ($nor$) and curvature ($curvature$) 
improves the prediction of the B-spline filling surface. In the ablation, we keep the architecture and training protocol identical and apply the same cross attention weighting mechanism to all input modalities;  $nor$ $curvature$
(when provided) are incorporated into the point tokens and participate in the cross-attention in the same way as $P$.
\begin{table}[t]
\centering
\caption{\textbf{Input modality ablation.} Ablation of input modalities for parameter regression in N-sided hole filling.}
\label{tab:input_ablation}
\begin{tabular}{lc}
\toprule
Input & MSE to target parameters \\
\midrule
$P$ & 5.84e-3 \\
$(P,\ \mathrm{nor})$ & 5.87e-3 \\
$(P,\ \mathrm{nor},\ \mathrm{curvature})$ & 5.79e-3 \\
\bottomrule
\end{tabular}
\end{table}
Table~\ref{tab:input_ablation} shows that additional differential attributes yield no measurable improvement in our setting. The differences are marginal and do not indicate a consistent improvement. We therefore conclude that, under our data and task setting, explicit differential attributes provide negligible additional information beyond 
$P$, while increasing preprocessing and input bandwidth; hence we use $P$ as the default input. Under smooth, noise-free boundary sampling, normals and curvature are largely redundant with coordinate geometry; the two-stage cross-attention refinement can implicitly recover such cues from $P$, which explains the negligible performance difference.

\textbf{Effectiveness of Voxel and Resolution.} Under identical boundary conditions, projection surfaces were generated using both voxel-based and coordinate-based methods, as shown in Figure~\ref{fig:vor}. The results obtained from the coordinate-based method exhibit significant deviation from the boundary conditions, making it unsuitable for effective parameterization. Based on the data presented in Figure~\ref{fig:vor} and Table~\ref{tab:pr}, we have validated the necessity of using voxel-based methods and confirmed the effectiveness of our progressive generation strategy.
\begin{table}[htbp]
\centering
\caption{\textbf{Component and backbone ablation.} Impact of different components and backbones in terms of parameter error.}
\label{tab:back}
\begin{tabular}{lcccc}
\toprule
 & Backbone & CA & PR & Parameter error \\
\midrule
 & \multirow{4}{*}{UVTran}  & $\times$ & $\times$ & 1.07e-2 \\
 &                          & $\checkmark$ & $\times$ & 7.55e-3 \\
 &                          & $\times$ & $\checkmark$ & 8.10e-3 \\
 &                          & $\checkmark$ & $\checkmark$ & \textbf{5.84e-3} \\
\midrule
 & \multirow{4}{*}{ResNet}  & $\times$ & $\times$ & 1.02e-2 \\
 &                          & $\checkmark$ & $\times$ & 9.25e-3 \\
 &                          & $\times$ & $\checkmark$ & 9.86e-3 \\
 &                          & $\checkmark$ & $\checkmark$ & 8.11e-3 \\
\midrule
 & \multirow{4}{*}{PTran}   & $\times$ & $\times$ & 4.36e-2 \\
 &                          & $\checkmark$ & $\times$ & 2.37e-2 \\
 &                          & $\times$ & $\checkmark$ & 2.98e-2 \\
 &                          & $\checkmark$ & $\checkmark$ & 9.25e-3 \\
\bottomrule
\end{tabular}
\end{table}

\textbf{Effects of each Component and Backbone.}
To further verify the effectiveness of cross-attention (CA), and the progressive-resolution (PR) training strategy, we tested the performance of each module across different backbones, including residual networks (ResNet) and Point Transformer (PTran)~\cite{zhao2021}. As shown in Table~\ref{tab:back}, after incorporating the cross-attention mechanism and progressive resolution strategy, the performance based on the traditional transformer outperforms the others, achieving the best results. This validates the necessity of each module and confirms the rationale for selecting the traditional transformer as the backbone.

\section{Conclusion}
This paper proposes a deep learning model, UVTran, which leverages the locality of B-splines and employs cross-attention to extract local features, generating reliable projection surfaces. The hole boundary is projected onto the projection surface to obtain the parametric curve, thereby improving surface quality. Experimental results on both test datasets and benchmarks demonstrate the effectiveness and robustness of the proposed model. Generating solutions that directly satisfy continuity constraints remains challenging. Therefore, the output of the model serves only as the projection surface, which is then used to generate the parametric curve. To overcome this limitation, future work will focus on expanding the dataset and employing model ensemble techniques to generate higher-precision solutions. Additionally, learning-based methods will be explored to address other fitting problems in industrial geometry design, such as curve and surface fitting, surface reconstruction, and shape optimization.

\appendix
\section{Appendix}
\label{sec:Appendix}
The following sections present detailed derivations of the formulas introduced in Section~\ref{sec:pre}.

When working with composite surfaces and their associated formulations, the nested summations and indices can be difficult to track. Welch proposed transforming the control-point matrix into a row vector.
\begin{equation}
    CP_{i,j} = \sum_k{S_{ijk}cp_{k}}
\end{equation}
where$S_{ijk}$ = 1 if $CP_{ij}$ corresponds to $p_{k}$, and $S_{ijk}=0$ otherwise.
\begin{equation}
\label{row}
    b_{k}(u, v)=\sum_{i} \sum_{j} N_{i}(u) N_{j}(v) S_{i j k}
\end{equation}

By defining the energy function $E_v$, the N-sided hole-filling problem is formulated as the following minimization problem:
\begin{equation}
\min  E_{v}=\min (E_{\text {Surface }}+E_{\text {Cons }})
\end{equation}
Using equation~\ref{row}, $E_{bending}$ can be written as

\begin{equation}
\begin{split}
    E_{bending}=
    \\
    \int\int \sum_{i, j=1}^{2}\left({c}
\alpha_{i j} \mathbf{(cp)}^{T} D_{i} \mathbf{b} \mathbf{(cp)}^{T} D_{j} \mathbf{b} 
+ 
\beta_{i j}\left(\mathbf{(cp)}^{T} D_{i} D_{j} \mathbf{b}\right)^{2}
\right)=
\\
\mathbf{(cp)}^{T} \int\int\sum_{i . j=1}^{2}\left(\begin{array}{c}
\alpha_{i j} D_{i} \mathbf{b} \otimes D_{j} \mathbf{b} \\
+ \\
\beta_{i j} D_{i} D_{j} \mathbf{b} \otimes D_{i} D_{j} \mathbf{b}
\end{array}\right) \mathbf{(cp)} \\
 =\mathbf{(cp)}^{T} \mathbf{A} \mathbf{(cp)}
\end{split}
\end{equation}

Here, $D_{j}\mathbf{b}$ denotes the corresponding partial derivatives. The same procedure applies to the computation of $E_{\mathrm{roc\ in\ bending}}$.

\begin{equation}
    E^{Cur}_{Position} 
 =\mathbf{(cp)}^{T} \mathbf{D} \mathbf{(cp)}-2\mathbf{b} \mathbf{(cp)}+\mathbf{b}^{2}
\label{g0}
\end{equation}

Similarly, the normal and $C^{2}$ continuity constraints can be written as Eq.~\ref{g0}, where $\mathbf{b}$ denotes the corresponding constraint term.
\begin{equation}
\min  E_{v} = \min (\mathbf{(CP)^T}(\mathbf{A}+\mathbf{D}) \mathbf{(CP)}-2 \mathbf{b} \mathbf{(CP)}+C)
\end{equation}
The matrices and vectors are defined as:
\begin{equation}
    \mathbf{A}=\mathbf{A}_{1}+\mathbf{A}_{2} 
\end{equation}
\begin{equation}
\mathbf{D}= \sum_{k=1}^{l_{0}+l_{1}+l_{2}+\cdots} \mathbf{D}_{1}^{l}+ \sum_{k=l_{0}+1}^{l_{0}+l_{1}+l_{2}+\cdots} \mathbf{D}_{2}^{l}+\sum_{l=l_{0}+l_{1}+1}^{l_{0}+l_{1}+l_{2}+\cdots} \mathbf{D}_{3}^{l} 
\end{equation}
\begin{equation}
\mathbf{b}= \sum_{k=1}^{l_{0}+l_{1}+l_{2}+\cdots} \mathbf{b}_{1}^{l}+ \sum_{k=l_{0}+1}^{l_{0}+l_{1}+l_{2}+\cdots} \mathbf{b}_{2}^{l}+\sum_{l=l_{0}+l_{1}+1}^{l_{0}+l_{1}+l_{2}+\cdots} \mathbf{b}_{3}^{l} 
\end{equation}
\begin{equation}
C=C_{1}+C_{2}+C_{3}
\end{equation}

Here, $\mathbf{CP}$ denotes the stacked vector of control points of the target surface. $\mathbf{A}$ is the coefficient matrix corresponding to the surface-energy term. $\mathbf{A}{1}$ and $\mathbf{A}{2}$ are the matrices corresponding to the $E_{roc in bending}$ and $E_{bending}$, respectively. $\mathbf{D}$ encodes the contributions induced by the surface terms evaluated along the boundary pcurve, and $\mathbf{b}$ collects the prescribed constraint terms. $l_{0}, l_{1}, l_{2}, \ldots$ denote the corresponding hole-boundary curves.

\bibliographystyle{cas-model2-names}

\bibliography{cas-refs}

\end{document}